\documentclass{nature}
\usepackage{graphicx}
\usepackage{lineno}
\usepackage[T1]{fontenc}
\pdfoutput=1

\newcommand{\iso}[2]{\hbox{${}^{#1}${#2}}}
\newcommand{\msun}{\ensuremath{{\rm M}_{\odot}}}

\newcommand{\apj}{Astrophys. J.}

\newcommand{\apjs}{Astrophys. J. Suppl.}
\newcommand{\aap}{Astron. Astrophys.}
\newcommand{\mnras}{Mon. Not. R. Astron. Soc.}
\newcommand{\araa}{Ann. Rev. Astron. Astrophys.}

\newcommand{\prc}{Phys. Rev. C}
\newcommand{\prl}{Phys. Rev. Lett.}

\title{Origin of meteoritic stardust unveiled by a new proton-capture 
rate of oxygen-17}

\author{M. Lugaro$^{1,2,*}$, A.~I. Karakas$^{2,3,4}$, C.~G. Bruno$^{5}$, M. 
Aliotta$^{5}$, L.~R. Nittler$^{6}$, D. Bemmerer$^{7}$,
A. Best$^{8}$, A. Boeltzig$^{9}$, C. Broggini$^{10}$, 
A. Caciolli$^{11}$, F. Cavanna$^{12}$, G.~F. Ciani$^{9}$,
P. Corvisiero$^{12}$, T. Davinson$^{5}$, R. Depalo$^{11}$, 
A. Di Leva$^{8}$, Z. Elekes$^{13}$, F. Ferraro$^{12}$, 
A. Formicola$^{14}$,
Zs. F\"ul\"op$^{13}$, 
G. Gervino$^{15}$, A. Guglielmetti$^{16}$, C. Gustavino$^{17}$
Gy. Gy\"urky$^{13}$, G. Imbriani$^{8}$, M. Junker$^{14}$,
R. Menegazzo$^{10}$, V. Mossa$^{18}$, F.~R. Pantaleo$^{18}$, D. Piatti$^{11}$,
P. Prati$^{12}$,
D.~A. Scott$^{5}$,
O. Straniero$^{19}$, F. Strieder$^{20,21}$, T. Sz\"ucs$^{13}$, M.~P. Tak\'acs$^{7}$,
D. Trezzi$^{16}$}

\begin{document}

\maketitle

\noindent $^*${\it Corresponding author}
\begin{affiliations}
 \item Konkoly Observatory, Research Centre for Astronomy and Earth Sciences, 
Hungarian Academy of Sciences, H-1121 Budapest, Hungary
 \item Monash Centre for Astrophysics (MoCA), Monash University, Clayton VIC
3800, Australia
 \item Research School of Astronomy and Astrophysics, Australian National
University, Canberra, ACT 2611, Australia
 \item
Kavli Institute for the Physics and Mathematics of the Universe (WPI), University of Tokyo, 
Kashiwa, Chiba 277-8583, Japan
 \item SUPA, School of Physics and Astronomy, University of Edinburgh, EH9 3FD 
Edinburgh, United Kingdom
 \item Department of Terrestrial Magnetism, Carnegie Institution for Science, 
Washington, DC 20015, USA
 \item Helmholtz-Zentrum Dresden-Rossendorf, Bautzner Landstr. 400, 01328 
Dresden, Germany
\item Universit\`a di Napoli Federico II and INFN, Sezione di Napoli, 80126 Napoli, Italy
\item Gran Sasso Science Institute, INFN, Viale F. Crispi 7, 67100 L'Aquila, Italy
\item INFN of Padova, Via Marzolo 8, I-35131 Padova, Italy
\item Universit\`a degli Studi di Padova and INFN, Sezione di Padova, Via F. Marzolo 8, 35131 Padova, Italy
\item Universit\`a degli Studi di Genova and INFN, Sezione di Genova, Via Dodecaneso 33, 16146 Genova, Italy
\item Institute for Nuclear Research (MTA ATOMKI), PO Box 51, HU-4001 Debrecen, Hungary
\item INFN, Laboratori Nazionali del Gran Sasso (LNGS), 67100 Assergi, Italy
\item Universit\`a degli Studi di Torino and INFN, Sezione di Torino, Via P. Giuria 1, 10125 Torino, Italy
\item Universit\`a degli Studi di Milano and INFN, Sezione di Milano, Via G. Celoria 16, 20133 Milano, Italy 
\item INFN, Sezione di Roma La Sapienza, Piazzale A. Moro 2, 00185 Roma, Italy
\item Universit\`a degli Studi di Bari and INFN, Sezione di Bari, 70125 Bari, Italy
\item Osservatorio Astronomico di Collurania, Teramo, and INFN, Sezione di Napoli, 80126 Napoli, Italy
\item Institut f\"ur Experimentalphysik III, Ruhr-Universit\"at Bochum, 44780 Bochum, Germany
\item Present address: South Dakota School of Mines, 501 E. Saint Joseph St., SD 57701 USA

\end{affiliations}

\begin{abstract}

Stardust grains recovered from meteorites provide high-precision snapshots of the isotopic 
composition of the stellar environment in which they formed\cite{zinner14}. Attributing their 
origin to specific types of stars, however, often proves difficult. Intermediate-mass stars 
of 4-8 solar masses are expected to contribute a large fraction of meteoritic 
stardust\cite{gail09,zhukovska15}. However, no grains have been found with characteristic 
isotopic compositions expected from such stars\cite{lugaro07,iliadis08}. This is a 
long-standing puzzle, which points to serious gaps in our understanding of the lifecycle of 
stars and dust in our Galaxy. Here we show that the increased proton-capture rate of
\iso{17}O reported by a recent underground experiment\cite{bruno16} leads to 
\iso{17}O/\iso{16}O isotopic ratios that match those observed in a population of stardust 
grains, for proton-burning temperatures of 60--80 million K. These temperatures are indeed 
achieved at the base of the convective envelope during the late evolution of 
intermediate-mass stars of 4-8 solar masses\cite{ventura13,cristallo15,karakas16}, which 
reveals them as the most likely site of origin of the grains. This result provides the first 
direct evidence that these stars contributed to the dust inventory from which the Solar 
System formed.

\end{abstract}


Stardust grains found in meteorites (and also interplanetary dust particles and
samples returned from comet Wild 2) represent the very small fraction 
of presolar dust that 
survived destruction in the protosolar nebula. They condensed in the atmospheres 
of evolved stars and in nova and supernova ejecta and were preserved inside 
meteorites\cite{zinner14}. Their isotopic compositions are measured with high 
precision (few percent uncertainties) via mass spectrometry and represent 
a direct record of their site of formation, providing us with deep  
insights into stellar physics and the origin of elements and of dust in the Galaxy. 
Identified stardust includes both 
carbon-rich (diamonds, graphite, 
silicon carbide) and oxygen-rich (e.g., Al-rich oxides, silicate) grains, with 
the former condensing from gas with C$>$O, and the latter from gas with C$<$O. 
Here we focus on oxide and silicate grains, which are  
classified into different groups 
mostly based on their oxygen isotopic compositions\cite{nittler97}.
Group I grains make up the majority ($\sim$75\%) 
of oxide and silicate grains and show excesses 
in \iso{17}O 
characteristic of the first dredge-up in red giant stars of initial mass roughly 1--3 
\msun, with a maximum \iso{17}O/\iso{16}O$\sim$0.003. 
Their origin is generally well understood and 
attributed to the O-rich phases of the subsequent
asymptotic giant branch (AGB), when large amounts of dust condense in the 
cool, expanding stellar envelopes\cite{gail09}. Group II grains represent roughly 10\% 
of all presolar oxide grains, although this is a lower limit since 
measured compositions may suffer
from isotopic dilution during ion probe analysis.
Like Group I grains they 
display excesses in \iso{17}O (with \iso{17}O/\iso{16}O up to 0.0015), 
but are also highly depleted in \iso{18}O, 
with \iso{18}O/\iso{16}O ratios 
down by two orders of magnitude relative to the solar value.
The initial ratio of the radioactive \iso{26}Al (half life, T$_{1/2} = 0.7$ Myr) to 
\iso{27}Al is inferred from \iso{26}Mg excesses and in Group II grains 
reaches 0.1, almost an order 
of magnitude higher than in Group I grains, on average. 
While this composition is the 
indisputable signature of H burning activating proton captures 
on the oxygen isotopes and on \iso{25}Mg
(the \iso{25}Mg(p,$\gamma$)\iso{26}Al reaction), hypotheses on the 
site of formation of Group II grains are still tentative.

Hydrogen burning affects the surface composition of massive ($>$ 4 \msun) AGB 
stars when the base of the convective envelope becomes hot 
enough for proton-capture nucleosynthesis to occur\cite{ventura13} 
(``hot bottom burning'', 
HBB, Figure~1). These are the brightest AGB stars, and the fact that they 
mostly show C/O$<$1 is attributed to the operation of the CN cycle, 
which depletes carbon\cite{wood83}, 
in contrast to their less bright counterparts, which mostly show C/O$>$1 
as a result of the dredge-up of He-burning material rich in carbon.
Characteristic temperatures of HBB exceed $\sim$60~MK and,
thanks to the fast convective turn-over time ($\approx 1$~yr), 
the composition of the whole envelope is quickly transmuted into the H-burning equilibrium 
abundances produced at the base of the envelope. Massive AGB stars are observed to generate 
significant amounts of dust and based on current models of Galactic dust evolution
are expected to have 
contributed almost half of the O-rich dust of AGB origin in the Solar 
System\cite{gail09,zhukovska15}. However, no stardust grains have been found to show 
the signature of HBB because, although Group II grains show 
the highly depleted \iso{18}O/\iso{16}O ratios qualitatively expected from HBB, 
their \iso{17}O/\iso{16}O ratios 
are roughly
two-times lower than predicted\cite{lugaro07,iliadis08} 
using the available reaction rates\cite{iliadis10}. 

Currently, the preferred
suggestion for the origin of Group II grains is that 
they formed in AGB stars of low mass ($<$1.5 \msun) that did not dredge-up 
enough carbon to become C-rich but experienced  
extra mixing below the bottom of the 
convective envelope (``cool bottom processing'', CBP\cite{nollett03,palmerini11}, Figure 1). 
In this scenario, material from 
the bottom of the convective envelope penetrates the thin radiative region 
located between the base of the convective envelope and the top of the 
H-burning shell where the temperature and density increase steeply with   
mass depth and proton captures can occur (Figure~1).
While mechanisms 
have been proposed to explain the physical process driving this extra 
mixing\cite{nucci14} the current modelling of the CBP is parametric: 
both the rate of the extra mixing and the depth reached are treated 
as a free parameters, with the depth adjusted in order to reach temperatures 
in the range 40--55 MK.  


Whichever scenario we consider, the equilibrium \iso{17}O/\iso{16}O ratio produced by H 
burning is determined by the competition between the processes that produce and 
destroy \iso{17}O. Specifically, it depends on the ratio between the rate of the 
\iso{16}O(p,$\gamma$)\iso{17}F reaction, which produces \iso{17}O following the beta decay 
of \iso{17}F (T$_{1/2} = 64$~s), and the rate of the \iso{17}O(p,$\alpha$)\iso{14}N 
reaction, which destroys \iso{17}O. (Note that the \iso{17}O(p,$\gamma$)\iso{18}F is 
comparatively negligible at all temperatures considered here). The former is known to within 
7\%\cite{iliadis08,iliadis10}; the latter has recently been determined\cite{bruno16} from a 
direct measurement of the strength of the 64.5~keV resonance that dominates the 
\iso{17}O(p,$\alpha$)\iso{14}N reaction rate at temperatures between 10 and 100 
MK\cite{bruno16}, i.e., over the entire range of interest here. The experiment took place at 
the Laboratory for Underground Nuclear Astrophysics (LUNA) at Gran Sasso, Italy, where 
improved experimental procedures and a 15-times 
lower background for $\alpha$ particles detection than in surface laboratories allowed 
for the most sensitive measurement to date\cite{bruno16}. The new rate is 2--2.5 times 
higher than previous evaluations\cite{iliadis10,buckner15}.  At the temperatures of 40--55 
MK typical of CBP the new rate reproduces only the lowest \iso{17}O/\iso{16}O values 
observed in Group II grains (Figure~2). At the temperatures of 60--80 MK 
typical of HBB, instead, the new rate reproduces most of the observed \iso{17}O/\iso{16}O 
range, revealing for the first time the signature expected from HBB in stardust grains. 
HBB temperatures higher than $\sim$ 80 MK are excluded for the parent stars of the grains.

Although the initial stellar mass and metallicity ranges at which HBB occurs as well as the AGB 
lifetime are model 
dependent\cite{ventura13,cristallo15,karakas16}, our result is robust because any massive AGB model 
experiencing HBB with temperatures between 60 and 80 MK will necessarily 
produce \iso{17}O/\iso{16}O ratios in agreement with 
those observed in most Group II grains. Figure~3 shows the surface evolution of 
the oxygen isotopic ratios 
for three AGB models (of initial mass 4.5, 5, and 6 \msun) of solar metallicity that experience HBB (see 
Methods section), compared to observed isotopic ratios in Group II stardust grains. The models evolve 
through the 
first and second dredge-ups at the end of core H- and He-burning, respectively, which increase the 
\iso{17}O/\iso{16}O ratio by roughly a factor of five. During the subsequent AGB phase, HBB quickly 
(e.g., after about 1/5 of its total TP-AGB lifetime for a 6\msun\ star)
shifts the oxygen 
isotopic composition to the equilibrium values corresponding to the burning temperature. Using the 
LUNA rate, the 
\iso{17}O/\iso{16}O ratio produced by HBB is roughly a factor of 2 lower than that obtained with the 
previous 
rate by Iliadis {\it et al.}\cite{iliadis10} and nicely reproduces those observed
 in Group II grains, 
when AGB material is diluted with material of solar composition.
The dilution is required because HBB strongly depletes \iso{18}O. This is 
in accordance with the 
non-detection of \iso{18}O in bright O-rich AGB 
stars\cite{justtanont15}, but results in \iso{18}O/\iso{16}O ratios more than two orders of magnitude 
lower than 
observed in Group II grains. Dilution with solar material is particularly effective
in increasing the \iso{18}O/\iso{16}O ratio: for example, 99\% of 
HBB material mixed with only 1\% of solar material increases the \iso{18}O/\iso{16}O by two orders of 
magnitudes. 
On the other hand, dilution has a comparatively minor effect on the other
isotopes measured in the grains because \iso{17}O, \iso{25}Mg, and \iso{26}Al are
{\em produced} rather than {\em destroyed} in massive AGB stars.
For example, it takes dilution with 50\% of solar system material to decrease
the \iso{17}O/\iso{16}O and \iso{25}Mg/\iso{24}Mg ratios by a factor of 2.

Dilution can be wrought by
percent-level traces of contaminant oxygen (e.g., 
from terrestrial or non-presolar material) during isotopic measurements, which can 
result in \iso{18}O/\iso{16}O up to 
$\sim$10$^{-4}$; however, laboratory contamination cannot easily explain grains with higher 
\iso{18}O/\iso{16}O values. For these, a dilution of the HBB signature composition with solar system 
material
at the level of a few tens of percent is required. 
Even higher dilution would result in a fraction of 
Group I 
grains also originating from massive AGB stars.
Possible processes may involve dilution with previously ejected gas within the 
dust formation region; 
dilution with material in the interstellar medium; and/or
a significantly lower value of the
\iso{18}O(p,$\alpha$)\iso{15}N reaction rate. A study of this 
reaction has recently been completed at LUNA and data analysis is in progress.

The other isotopic pairs measured in Group II grains are also
consistent with an origin in massive AGB stars. The \iso{25}Mg/\iso{24}Mg ratios 
are enhanced in massive AGB stars by the third dredge-up of material from the He inter-shell, 
where the \iso{22}Ne($\alpha$,n)\iso{25}Mg reaction is 
activated, and such a signature is seen in some presolar spinel (MgAl$_{2}$O$_{4}$) grains 
(Figure~4a). 
Specifically, the two times solar \iso{25}Mg/\iso{24}Mg value observed in a spinel grain named 
14-12-7\cite{gyngard10} is 
close to that obtained in the final composition of the 5 \msun, model. 
However, grain OC2\cite{lugaro07} and the majority of the other grains show a spread in 
the \iso{25}Mg/\iso{24}Mg ratio from 1 to 1.5 times solar, i.e., 
lower than predicted by the dilution computed using  
the final AGB composition. This may reflect partial equilibration of Mg isotopes in 
the grains themselves\cite{nittler08}. 
Alternatively, the lower \iso{25}Mg/\iso{24}Mg ratios may be 
explained by truncating 
the AGB evolution to one-half or one-third of the total computed evolution 
(as illustrated in Figures~3 and 4). This could result from a higher mass-loss rate and/or the effect of binary 
interactions. Another solution allowed within current model uncertainties is 
a less efficient third dredge-up than calculated in our models.
Finally, the high \iso{26}Al/\iso{27}Al ratios up to $\sim$0.1 typical of 
Group II grains are also consistent with HBB (Figure~4b),
although an accurate analysis is currently hampered by the uncertainties in the 
\iso{25}Mg and \iso{26}Al proton-capture rates\cite{iliadis10,straniero13}. 

Our evidence that some meteoritic stardust grains exist whose O, Mg, and Al 
isotopic composition is best accounted for by H-burning conditions characteristic of massive 
AGB stars proves that these stars were dust contributors to the early Solar System. It 
further provides us with a new tool to deepen our understanding of uncertain physical 
processes in massive AGB stars, for which observational constraints are still scarce.





{\bf Acknowledgements} We thank Onno Pols and Rob Izzard for useful insights on binary systems and Paola Marigo
for discussion of our results. 
M.~L. is a Momentum (``Lend\"ulet-2014'' Programme) project leader of the Hungarian 
Academy of Sciences. M.~L. and A.~I.~K. are grateful for the support of the NCI National Facility 
at the ANU.

{\bf Author contributions} M.~L. designed and carried out the research, run the nucleosynthesis models, 
prepared the figues, and wrote the paper. A.~I.~K. run the stellar structure models, 
discussed the results, 
and wrote the paper. C.~G.~B. played a key role in the set up and running of the underground experiment 
of the $^{17}$O(p,$\alpha$)$^{14}$N reaction and analysed the data to derive the new rate. 
M.~A. contributed to run 
the experiment and wrote the paper. L.~R.~N. contributed to the collection of the stardust grain data, 
discussed the results, prepared the figures, and wrote the paper. The other authors are 
co-investigators who 
set up and ran the underground experiment that lasted about three years, from 2012 to 2015, and made the 
measurements possible. O.~S. also discussed the results. 

{\bf Author information}
Correspondence and requests for materials
should be addressed to M. Lugaro~(email: maria.lugaro@csfk.mta.hu).


\begin{figure}
\center{\includegraphics[scale=0.5]{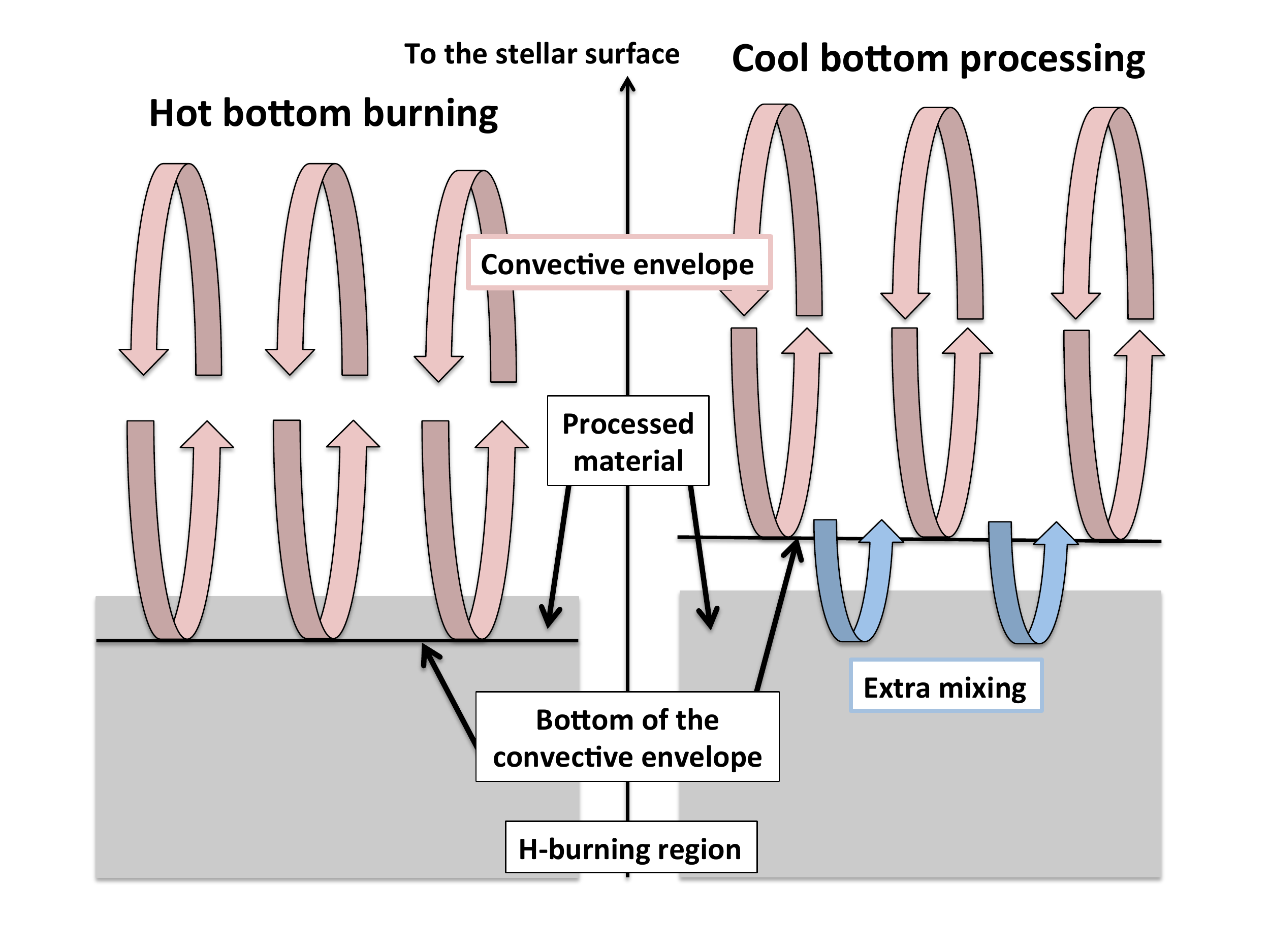}} 
\caption{Schematic diagram of the internal structure of AGB stars at the interface between the 
H-burning region and the convective envelope. Hot bottom burning (HBB, left) and cool 
bottom processing (CBP, right) take place in massive and low-mass AGB stars, 
respectively, and carry material processed in the H-burning region to stellar surface. The main 
differences between the two cases are that: (1) material is processed at higher 
temperatures but lower densities in the case of HBB, with respect to CBP; and (2) 
mixing occurs via convection in the case of HBB, while 
non-convective extra mixing needs to be invoked in the case of CBP.
\label{fig:scheme}}
\end{figure}

\begin{figure}
\center{\includegraphics[scale=0.7]{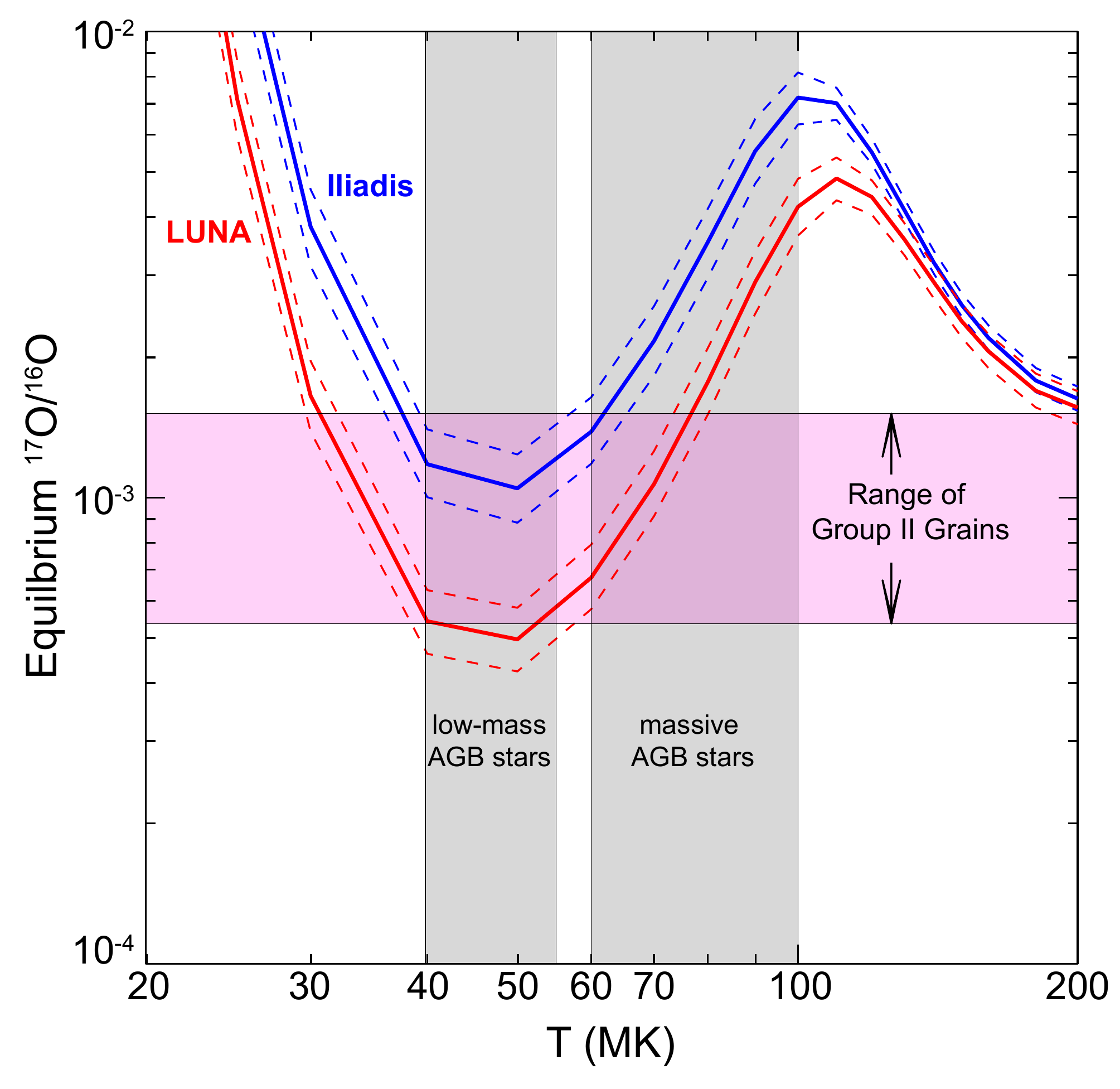}} 
\caption{Equilibrium \iso{17}O/\iso{16}O ratio defined as the 
ratio of the production to destruction rates of \iso{17}O in the temperature range of
interest for AGB stars. We used the recommended (thick solid lines)
and the lower and
upper limits (thin dashed lines, essentially corresponding to the 1$\sigma$
experimental uncertainty of the
strength of the 64.5~keV resonance) of the \iso{17}O(p,$\alpha$)\iso{14}N reaction rate from  
LUNA\cite{bruno16} and Iliadis {\it et al.}\cite{iliadis10}. 
The horizontal pink band 
shows the range of \iso{17}O/\iso{16}O values observed in Group II grains.
The typical temperature ranges for CBP 
in low-mass AGB stars and for HBB in massive AGB stars  
are shown as grey vertical bands.
\label{fig:ratiorates}}
\end{figure}

\begin{figure} 
\center{\includegraphics[scale=0.8]{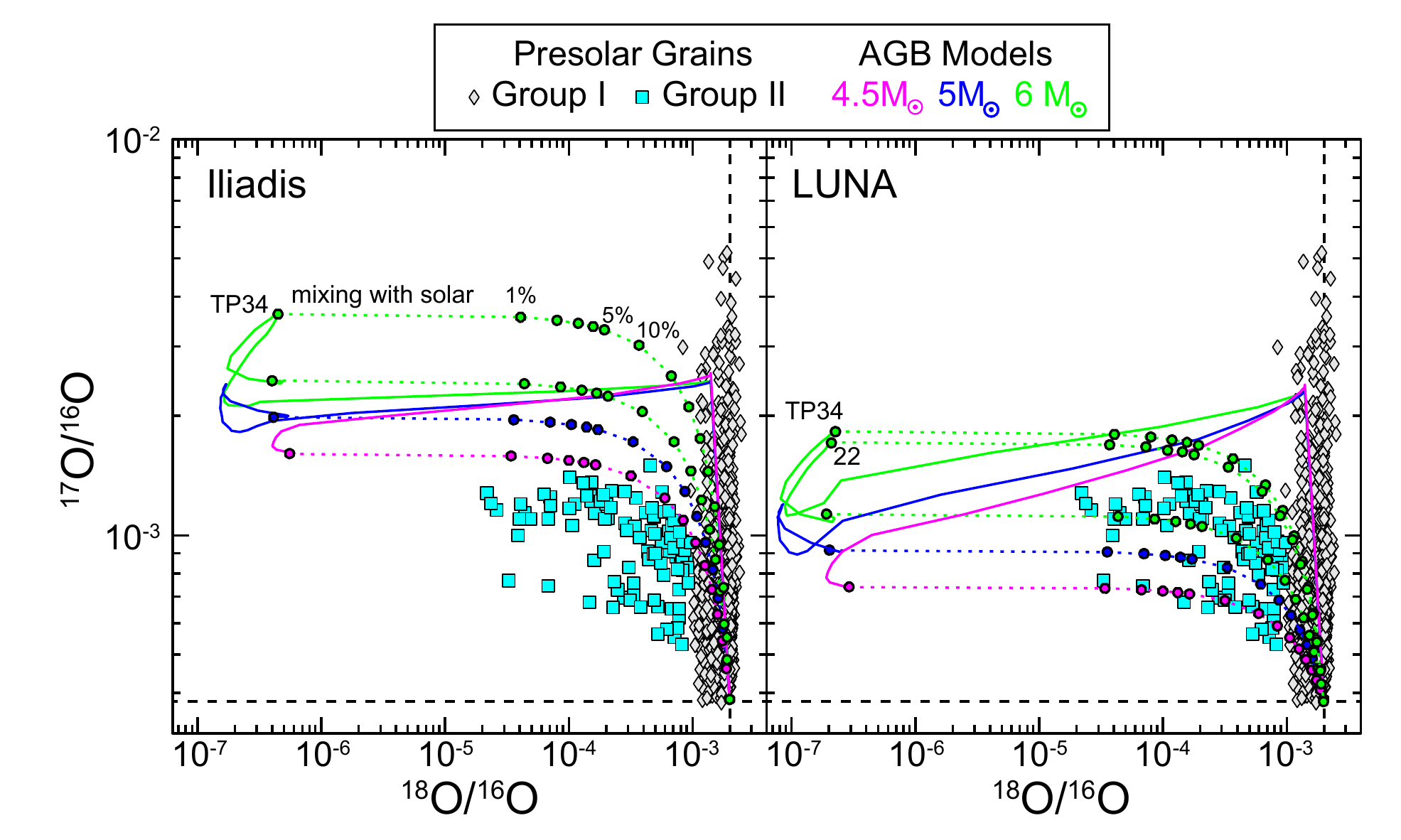}} 
\caption{
Evolution of the oxygen isotopic ratios at the surface of AGB models of different masses. 
The evolutionary (solid)  
lines in panel a were calculated using the old (Iliadis\cite{iliadis10}) and in panel b 
using the new 
(LUNA\cite{bruno16}) \iso{17}O(p,$\alpha$)\iso{14}N reaction rate. Uncertainties in 
either rate translate into changes in the \iso{17}O/\iso{16}O ratio 
by at most 20\%, i.e., within the differences 
between the different stellar models. Isotopic ratios observed in Group II grains (filled square 
symbols\cite{hynes09} with error bars typically within the size of the symbol) 
cannot be reproduced by the old rate, regardless of the amount of dilution of AGB material 
with solar material (dotted lines), but are well reproduced with the new rate.  
The dilution is applied to the AGB composition at the end of the 
evolution for the three masses and, as examples, also   
at one-half and one-third of the AGB lifetime for the 6 \msun\ star (labels TP34 
and 22 indicate that the star evolved, respectively, through 34 and 22 thermal instabilities 
of the He shell out of the 53 
computed in the models). Dashed vertical and horizontal lines indicate solar ratios for 
reference.
\label{fig:IliadisLUNA}}
\end{figure}

\begin{figure}
\center{\includegraphics[scale=0.8]{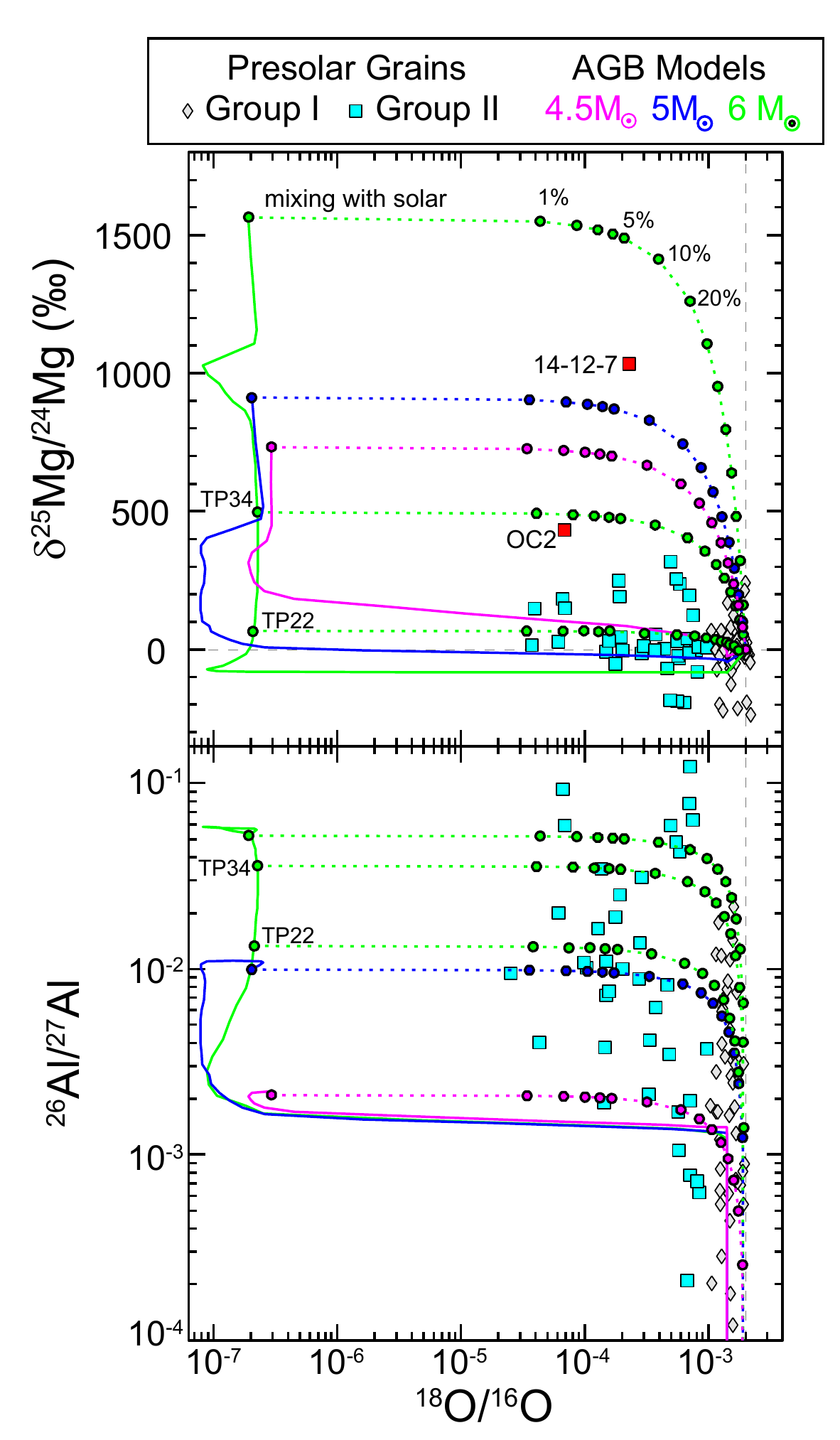}} 
\caption{Evolution of selected Mg versus O (a) and Al versus O (b) 
isotopic ratios at the surface of AGB models of different masses.
Same as Figure~3, with models calculated with the LUNA rate.  
$\delta$\iso{25}Mg/\iso{24}Mg-values represent permil variations with respect to the solar
system value. The two spinel grains with excesses in \iso{25}Mg
(OC2 and 14-12-7) are highlighted in red.
\label{fig:mg25mg26}}
\end{figure}

\newpage
\begin{methods}

\section*{Stellar models}
Stellar structure models 
with metallicities (Z) from half to double solar (where solar is 
0.014$^{\mbox{\scriptsize 22\if@tempswa , 1\fi}}$
were selected from the large set presented by 
Karakas$^{\mbox{\scriptsize 23\if@tempswa , 1\fi}}$
computed with the Monash-Stromlo code$^{\mbox{\scriptsize 24\if@tempswa , 1\fi}}$.
No mass-loss was assumed on the red giant branch and 
the Vassiliadis \& Wood$^{\mbox{\scriptsize 25\if@tempswa , 1\fi}}$
mass-loss formulation was used on the AGB. 
The C-rich and N-rich 
low-temperature opacity tables were taken from Marigo \& 
Aringer$^{\mbox{\scriptsize 26\if@tempswa , 1\fi}}$.
Convection was approximated using the mixing length theory with a mixing-length parameter of 1.86 
in all calculations. No 
convective overshoot was applied, although the algorithm described by 
Lattanzio$^{\mbox{\scriptsize 27\if@tempswa , 1\fi}}$
was used to search for a neutrally stable point for the border between convective and radiative 
zones. 

From the models presented by Karakas$^{\mbox{\scriptsize 23\if@tempswa , 1\fi}}$,
in the Supplementary Table~1
we present a selection with 
initial masses from 4.5 to 8 \msun\ with canonical values of the He content. 
More details on the physical quantities of the models can be found 
in Table~1 of Karakas$^{\mbox{\scriptsize 23\if@tempswa , 1\fi}}$.
In Supplementary Table~1 we only report a summary of 
those that are most relevant here: the total number of thermal instabilities of the He-burning shell 
(thermal pulses, TPs); 
the maximum temperature at the base of the 
convective envelope (${\rm T}^{\rm max}_{\rm bce}$); the maximum temperature achieved in the inter-shell 
(${\rm T}^{\rm max}_{\rm inter-shell}$); and 
the mass lost during the whole evolution (${\rm M}_{\rm lost}^{\rm total}$).
All the models experienced ${\rm T}^{\rm max}_{\rm bce}$ high 
enough to activate HBB, except for the 5 \msun\, model of Z$=0.03$.
It should be noted that the mass and
metallicity ranges at which HBB occurs are model dependent: for the same mass and metallicity,
models using more or less efficient convection, for example, via a different 
mixing length parameter or different mixing schemes, result in different 
temperatures\cite{ventura13,cristallo15}.
All our stellar models also experienced efficient third dredge-up, i.e., 
C-rich material being carried from 
the He-rich inter-shell to the convective envelope. This is also model
dependent.  
Overshooting at the base of the convective region associated with the TPs
is not included in our models. In combination with the 
third dredge-up it would enrich the envelope in \iso{16}O. However, 
this would not change the oxygen isotopic ratios at the stellar surface since 
HBB efficiently brings them  
to their equilibrium values, similarly to the case of the carbon isotopic ratios. 
In the lower-mass stars where HBB is not activated, efficient third dredge-up of 
\iso{16}O would be accompanied by efficient third dredge-up of 
\iso{12}C, producing a C-rich envelope where the oxide and silicate grains considered here 
do not form.

Because of both the large dilution and the effect of HBB, most of the models  
lead to O-rich surfaces -- the 
condition for the formation of oxide and silicate grains of interest here -- during their whole evolution, 
except for the 4.5 \msun\, model of Z=0.014 and the
5 \msun\, model of Z=0.007. These latter become C-rich after the 
second last TP and the last TP, respectively, which results in 40--50\% of the 
material ejected to be C-rich. For all the models, 
a relatively large fraction of the envelope material 
(20--30\%) is still present when our calculations stopped 
converging. The abundances we calculated for the last model are either lower
limits or a good approximation to the final enrichment, depending on 
possible further occurrence 
of third dredge-ups episodes beyond the point where our models stop converging.

We fed the computed stellar structure into the Monash post-processing code to calculate the detailed 
nucleosynthesis by solving simultaneously the abundance changes wrought by nuclear 
reactions and by convection using a ``donor cell'' advective scheme with two-streams (up and down) 
mixing. The simultaneous treatment of mixing and burning is required 
to model HBB in detail because nuclear reactions occur that may have timescales similar or shorter 
than the 
mixing timescales, also as function of the location in the envelope.
In these cases it is not possible to make the assumption of istantaneous mixing at an average 
burning rate. Essentially our method couples mixing and burning together in the post processing to 
obtain the nucleosynthesis, while the energetic feedback of HBB is 
taken from the structure calculations performed using instantaneous mixing.
The nucleosynthesis of elements up to Pb and Bi from the complete set of 
the models of Karakas$^{\mbox{\scriptsize 23\if@tempswa , 1\fi}}$
with He canonical abundance can be found in 
Karakas \& Lugaro\cite{karakas16}, together with a full discussion of the results. Briefly, in 
models that experience HBB, ${\rm T}^{\rm max}_{\rm bce}$ is the main feature controlling the 
composition of the stellar surface, and specifically the oxygen and aluminium ratios that are 
measured in oxide and silicate stardust grains. In massive AGB stars of 
metallicity around solar the Mg composition is affected mainly  
by the activation of the 
\iso{22}Ne($\alpha$,n)\iso{25}Mg and \iso{22}Ne($\alpha$,$\gamma$)\iso{26}Mg reactions 
in the He-rich inter-shell -- 
where ${\rm T}^{\rm max}_{\rm inter-shell}$ is well above the activation temperature (T $\approx 300$~MK) 
of these reactions in all the models --
and the subsequent third dredge-up of this material to the 
stellar surface. With respect to Karakas \& Lugaro\cite{karakas16}, 
we updated the \iso{22}Ne+$\alpha$ reaction rates from 
Iliadis {\it et al.}\cite{iliadis10} to Longland 
{\it et al.}$^{\mbox{\scriptsize 28\if@tempswa , 1\fi}}$
and 
the \iso{25}Mg+$\gamma$ reaction rates 
from Iliadis {\it et al.}\cite{iliadis10} to Straniero {\it et al.}\cite{straniero13}. 
Also, in the present study, we limited 
our calculations to a small network of 77 nuclear species, from neutrons to 
sulfur, plus the elements around the Fe peak, as described, e.g., in 
Karakas$^{\mbox{\scriptsize 24\if@tempswa , 1\fi}}$.
This 
choice allowed us to 
run each model in a few hours and test different values of the 
\iso{17}O(p,$\alpha$)\iso{14}N reaction rate: the recommended, the upper limit, and 
the lower limit from both Iliadis {\it et al.}\cite{iliadis10} and LUNA\cite{bruno16}. 
For the \iso{16}O(p,$\gamma$)\iso{17}F rate we used 
the value recommended by Iliadis {\it et al.}\cite{iliadis10}, which has an uncertainty
of 7\%\cite{iliadis08}. 
For the initial composition, we used Asplund 
{\it et al.}$^{\mbox{\scriptsize 22\if@tempswa , 1\fi}}$
for the solar metallicity models,
scaled down or up by factor of two for the Z=0.007 and Z=0.03 models, respectively. 
While we calculated detailed model 
predictions for all the models listed in the Supplementary Table~1, in the 
figures and discussion we focus on the 4.5, 5, and 6 \msun\, models with Z=0.014 only, for sake of 
clarity. 
Models with different metallicities
in the same mass range have similar ${\rm T}^{\rm max}_{\rm bce}$ and 
provide similar results, except for the 5 \msun model of Z=0.03, which does not experience HBB 
but remains O-rich due to the low efficiency of the third dredge-up combined with the high initial 
O abundance.
On the other hand, the 8 \msun\, models have 
too high ${\rm T}^{\rm max}_{\rm bce}$ to provide a match with the grain data 
(see Figure~2).
Stellar population synthesis models are needed to assess whether 
a number of Group I grains may also 
have originated from super-solar metallicity O-rich massive AGB stars that do not experience HBB.

\end{methods}

\begin{table*}[h]
\caption{Summary of the physical properties of a selection of AGB models with
initial masses (M, Column 1) from 4.5 to 8 \msun\ and metallicities
(Z, Column 2) half-solar (0.007), solar (0.014), and twice-solar (0.03). 
Column 3: the total number of thermal pulses (TPs) of the He-burning shell;
Column 4: the maximum temperature at the base of the
convective envelope; Column 5: the maximum temperature in the inter-shell;
Column 6: the mass lost during the whole evolution.}
\label{tab:models}
\begin{center}
\vspace{0.5cm}
\begin{tabular}{lccccc} \hline
\hline
M (\msun) & Z & No. of TP & ${\rm T}^{\rm max}_{\rm bce}$ (MK) & 
${\rm T}^{\rm max}_{\rm inter-shell}$ (MK) & 
${\rm M}_{\rm lost}^{\rm total}$ (\msun) \\
5 & 0.007 & 59 & 82.9 & 357 & 4.11 \\
4.5 & 0.014 & 31 & 63.5 & 356 & 3.64 \\
5 & 0.014 & 41 & 75.4 & 354 & 4.12 \\
6 & 0.014 & 53 & 85.5 & 365 & 5.08 \\
8 & 0.014 & 67 & 100 & 376 & 6.94 \\
5 & 0.03 & 26 & 54.2 & 345 & 4.13 \\
5.5 & 0.03 & 31 & 64.7 & 348 & 4.62 \\
6 & 0.03 & 33 & 71.2 & 352 & 5.09 \\
8 & 0.03 & 63 & 94.0 & 373 & 6.95 \\
\hline \hline
\end{tabular}
\end{center}
\end{table*}

\end{document}